\documentclass{ws-ijmpe}
\usepackage[super,compress]{cite}
\usepackage{amsmath}
\usepackage{amssymb}
\usepackage{graphicx}

\def\Lag{{\cal L}} 
\def\expect#1{{\left\langle #1 \right\rangle}}

\title{Effects of Tsallis distribution on parametric resonance in chiral phase transitions}
\author{Masamichi Ishihara}

\begin{document}
\maketitle

\begin{abstract}
The parametric resonance was studied in chiral phase transitions 
when the momentum distribution is described by a Tsallis distribution.
A Tsallis distribution has two parameters, the temperature $T$ and the entropic index $q$.
The amplification was estimated in two cases: 
1) expansionless case and 2) one dimensional expansion case. 
In an expansionless case, the temperature $T$ is constant,    
and the amplified modes as a function of $T$ were calculated for various $q$.
In one dimensional expansion case, the temperature $T$ decreases as a function of the proper time, 
and the amplification as a function of the transverse momentum was calculated for various $q$. 
In the expansionless case, the following facts were found:
1) the larger the value $q$ is, the softer the amplified modes are for the first and second resonance bands, 
2) the amplified mode of the first resonance band decreases and vanishes, as the temperature $T$ increases, 
and 
3) the amplified mode of the second resonance band decreases and approaches to zero, as the temperature $T$ increases. 
In one dimensional expansion case, the following facts were found:
1) the soft mode is amplified,  
2) the amplification is extremely strong around the amplified mode of the first resonance band at $T=0$, 
and 
3) the magnitude of the amplification as a function of transverse momentum oscillates around the amplified mode of the first resonance band at $T=0$. 
\end{abstract}

\ccode{25.75.Nq, 12.40.-y, 11.30.Rd, 25.75.-q}

\section{Introduction}
A power-like distribution appears in many branches of science. 
The Tsallis distribution is one of power-like distributions, and has been studied recent few decades. 
The distribution is an extension of the Boltzmann-Gibbs distribution.
A Tsallis distribution has two parameters: the temperature $T$ and the entropic parameter $q$.
The distribution has been applied to various phenomena \cite{TsallisBook}, 
and an example is momentum distribution at high energy collisions
\cite{Alberico2000,Biyajima2005,Biyajima2006,Wilk07_Tsallis,Wilk2009,Cleymans2012,Marques2015}.
 
In high energy heavy ion collisions, the phase transition is an important phenomena. 
Distribution affects the phase transition. 
The equation of state in the Tsallis nonextensive statistics was studied \cite{Drago2004,Pereira2007}, 
and the Nambu and Jona-Lasinio model was used in the study of the phase transition \cite{Rozynek2009}. 
The linear sigma model was used to study the effects of the distribution \cite{Ishihara2015}.
It was shown in these studies that  
the distribution affects physical quantities such as critical temperature, mass, etc. 

The enhancement of the field by parametric resonance was studied in the chiral phase transition
\cite{Hiro-Oka-1998, Hiro-Oka-1998-erratum,Ishihara2000,Ishihara2001}.
The condensate moves periodically and the soft mode is enhanced by the motion of the condensate. 
The parametric resonance occurs if the temperature is approximately constant, 
because the condensate shows approximate periodic motion.
Therefore, the parametric resonance may occur in expansionless and one dimensional expansion cases, 
even when the momentum distribution is described by a Tsallis distribution. 
It was pointed out that the momentum distribution at high energies is fitted well by a Tsallis distribution. 
Therefore, the effects of the Tsallis distribution on parametric resonance should be studied in the chiral phase transitions.

It is expected that the amplified mode by the parametric resonance is affected by the distribution.
The amplified mode by the parametric resonance is determined by the mass, 
because the mode is related to the oscillation of the condensate. 
The mass is related to the fluctuation of the field which is affected by the distribution.
Therefore, the distribution affects the amplified mode.

The purpose of this paper is to clarify the effects of the Tsallis distribution on the parametric resonance in the chiral phase transition,
when the momentum distribution is described by a Tsallis distribution. 
The amplified modes were obtained in an expansionless case, 
and the magnitude of the amplification was calculated in one dimensional expansion case. 
The parameter dependences of the amplified modes were studied to make the effects of the distribution clear.   

The following facts were found for the pion field.
In the expnasionless cases, these facts are
1) the larger the value $q$ is, the softer the amplified modes are for the first and second resonance bands, 
2) the amplified mode of the first resonance band decreases and vanishes, as $T$ increases, 
and 
3) the amplified mode of the second resonance band decreases and approaches to zero, as $T$ increases. 
In one dimensional expansion cases,  these facts are
1) the soft mode is amplified,  
2) the amplification is extremely strong around the amplified mode of the first resonance band at $T=0$, 
and 
3) the magnitude of the amplification as a function of the transverse momentum oscillates around the amplified mode of the first resonance band at $T=0$.

This paper is organized as follows. 
In section~\ref{sec:eq_of_motion_for_soft_mode}, 
the equations of soft modes are derived in expansionless and one dimensional expansion cases. 
The equations of the soft modes are derived in the linear sigma model when the condensate moves periodically.   
In section~\ref{sec:amplified_modes},
the amplified modes are studied. 
The amplified modes are derived analytically from the derived equations in the expansionless case,  
and $T$ and $q$ dependences of the amplified modes are shown.
The amplification for soft modes are studied numerically in one dimensional expansion case,   
and the amplification as a function of the transverse momentum is shown for some values of $q$. 
Section~\ref{sec:discussion_and_conclusion} is assigned for the discussion and conclusion.


\section{Equation of motion for soft mode}
\label{sec:eq_of_motion_for_soft_mode}
\subsection{Derivation of the equation for soft mode}
The Lagrangian of the linear sigma model is given by 
\begin{align} 
 \Lag = \frac{1}{2} \partial_{\mu} \phi  \partial^{\mu} \phi - \frac{\lambda}{4} \left( \phi^2 - v^2 \right)^2 + H \phi_0 ,
\end{align}
where $\phi$ represents $N$ scalar fields, $\phi \equiv (\phi_0, \phi_1, \cdots, \phi_{N-1})$.
The quantities $\phi^2$ and $\partial_\mu \phi \partial^\mu \phi$ represent 
${\displaystyle \phi ^2 \equiv \sum_{i=0}^{N-1} \left( \phi_i \right)^2}$ and 
${\displaystyle \partial_{\mu}  \phi \partial^{\mu}  \phi \equiv \sum_{i=0}^{N-1} \partial_{\mu}  \phi_i \partial^{\mu}  \phi_i}$, respectively. 

The field $\phi_i$ is divided into three parts, the condensate, soft modes, and  hard modes: 
\begin{equation}
\phi_i = \phi_{ic} + \phi_{is} + \phi_{ih} .
\label{eqn:div} 
\end{equation}
The statistical averages, $\expect{\phi_{ih}}$ and $\expect{\phi_{ih}\left(\phi_{jh}\right)^2}$,  are zero
when the free particle approximation is applied. 
The average $K_q(T) :=  \expect{\left( \phi_{ih} \right)^2}$ is independent of the suffix $i$ 
when the massless free particle approximation (MFPA) \cite{Gavin1994,Ishihara1999} is applied. 
In the present study, the statistical averages with respect to $\phi_{ih}$ are evaluated under MFPA.
The quantity $K_q(T)$ is given by the following integral under MFPA, when the distribution function is a Tsallis distribution $f_q(\vec{k})$:
\begin{equation}
K_q(T) = \int \ \frac{d\vec{k}}{(2\pi)^3} \ \frac{f_q(\vec{k})}{k} ,
\qquad f_{q}(\vec{k}) = \frac{1}{\left[1+ (q-1) \left(\frac{k}{T} \right)\right]_{+}^{\frac{1}{(q-1)}} - 1 } ,
\end{equation}
where $[x]_{+} = x$ for $x \ge 0$ and $[x]_{+} = 0$ for $x < 0$, and $k = | \vec{k} |$.
The quantity $K_q(T)$ is  represented \cite{Ishihara2015} with the digamma function $\psi(x)$ \cite{Abramowitz, Iwanami_III}:
\begin{equation}
K_q(T) = \frac{T^2}{2\pi^2 (q-1)} \left[ \psi(2-q) - \psi(3-2q) \right] \qquad \left( q < 3/2 \right) .
\end{equation} 

The averaged Lagrangian with respect to the hard modes is obtained 
by substituting Eq.~\eqref{eqn:div} into the Lagrangian and taking the statistical average.
The Lagrangian after this procedure is given by 
\begin{align}
\expect{\Lag}  = & 
\frac{1}{2} \partial_{\mu} \left( \phi_{c}+ \phi_{s} \right) \partial^{\mu} \left( \phi_{c}+ \phi_{s} \right)
    - \frac{\lambda}{4} \left[ \left( \phi_{c}+ \phi_{s} \right)^2 + (N+2) K_q(T) - v^2\right]^2 
\nonumber \\ & 
+ H \left( \phi_{0c}+ \phi_{0s} \right) + R, 
\label{averaged_Lagrangian}
\end{align}
where the term $R$ represents the terms that are independent of $\phi_{ic}$ and $\phi_{is}$. 

We define the effective potential $V(\phi_c)$ and the mass $m_j$ as follows:
\begin{equation}  
V(\phi_c) := \frac{\lambda}{4} \left[ \left( \phi_{c} \right)^2 + (N+2) K_q(T) - v^2\right]^2  - H \phi_{0c} ,
\end{equation}  
and 
\begin{equation}  
m_j^2  := \frac{\partial^2 V(\phi_c)}{\partial \phi_{jc}^2} 
              =  \lambda \left[ \left(\phi_c\right)^2 + 2 \left( \phi_{jc}\right)^2 + (N+2) K_q(T) - v^2 \right] , 
\end{equation}  
where ${\displaystyle \left(\phi_c\right)^2 = \sum_{i=0}^{N-1} \left( \phi_{ic} \right)^2}$.
The value of the condensate $\bar{\phi}_{jc}$ is defined at the minimum of the potential. 
Therefore, the condensate on the vacuum satisfies the following equation:
\begin{equation}
\left. \frac{\partial V(\phi_c)}{\partial \phi_{jc}}  \right|_{\phi=\bar{\phi}} 
= \lambda \left( \bar{\phi}_c^2 + (N+2) K_q(T) - v^2 \right) \bar{\phi}_{jc} - H \delta_{j0} 
= 0 .
\label{eqn:vev}
\end{equation}
The mass $m_j$ on the vacuum is represented as $\bar{m}_j$ hereafter. 

The Eular-Lagrange equation of $\phi_{jc}$ is derived from Eq.~\eqref{averaged_Lagrangian}: 
\begin{equation}
\Box \left( \phi_{jc} + \phi_{js} \right) 
+ \lambda \left[ \left( \phi_c + \phi_s\right)^2 + (N+2) K_q(T) - v^2 \right] \left( \phi_{jc} + \phi_{js} \right) 
- H \delta_{j0} = 0 .
\label{basic_eq_of_motion}
\end{equation}
The Eular-Lagrange equation of $\phi_{js}$ is the same equation. 
The lowest order equation of $\phi_{jc}$ is obtained by omitting the field $\phi_{js}$ from Eq.~\eqref{basic_eq_of_motion}: 
\begin{equation}
\Box \phi_{jc} + \lambda \left[ \left( \phi_c \right)^2 + (N+2) K_q(T) - v^2 \right]  \phi_{jc}  - H \delta_{j0} = 0 .
\label{basic_eq_of_motion:0th}
\end{equation}
The equation for $\phi_{js}$ is also obtained from Eq.~\eqref{basic_eq_of_motion}.
We take Eq.~\eqref{basic_eq_of_motion:0th} into account and  ignore $O\left( \phi_s^2 \right)$ terms. 
The equation is given by 
\begin{equation}
\Box \phi_{js} + \lambda \left[ \left( \phi_c \right)^2  + 2 \left( \phi_{jc} \right)^2  + (N+2) K_q(T) - v^2 \right] \phi_{js} 
+ 2 \lambda \sum_{\substack{i =0 \\ i \neq j}}^{N-1} \left( \phi_{ic}  \phi_{jc} \right) \phi_{is} = 0 
. 
\end{equation}
We set $\phi_{jc} = 0$ for $j \neq 0$, because  the potential is tilted to the $\phi_{0c}$ direction. 
The equations are reduced to the following equations:
\begin{subequations}
\begin{align}
& \Box \phi_{0c} + \lambda \left[ \left( \phi_{0c} \right)^2 + (N+2) K_q(T) - v^2 \right]  \phi_{0c}  - H  = 0 , 
\label{eqn:linearlized_equations:phic} \\
& \phi_{jc} = 0 \qquad (j \neq 0) ,\\
& \Box \phi_{js} + \lambda \left[ \left( 1 + 2 \delta_{j0} \right) \left( \phi_{0c} \right)^2  + (N+2) K_q(T) - v^2 \right] \phi_{js}  = 0 .
\label{eqn:linearlized_equations:phis} 
\end{align}
\label{eqn:linearlized_equations}
\end{subequations}
In the next subsection, we reconsider the above equations around the vacuum in specific cases.

\subsection{Equation for soft mode in an expansionless case}
In this subsection, we deal with the case that the temperature is constant.  
The starting point is Eq.~\eqref{eqn:linearlized_equations:phic},  
and the motion of $\phi_{jc}$ around the vacuum is derived.

The condensate $\bar{\phi}_{0c}$ satisfies the following equation from Eq.~\eqref{eqn:vev}:
\begin{equation}
\lambda \left[ \left(\bar{\phi}_{0c}\right)^2 + (N+2) K_q(T) - v^2 \right] \bar{\phi}_{0c} - H  = 0 ,
\end{equation}
and $\bar{\phi}_{jc}=0$ for $j \neq 0$. 
We define the field $\delta \phi_{0c}$ as  $\phi_{0c} = \bar{\phi}_{0c} + \delta \phi_{0c}$
and note that the condensate is independent of the coordinates in space. 
Therefore, Eqs.~\eqref{eqn:linearlized_equations:phic} and \eqref{eqn:linearlized_equations:phis} 
are rewritten with the field $\delta \phi_{0c}$:
\begin{subequations}
\begin{align}
& 
\left[ \frac{\partial^2}{\partial t^2} + \left( \bar{m}_0(T) \right)^2 \right] \left( \delta \phi_{0c} \right)  
+ O\left(\left( \delta \phi_{0c} \right)^2 \right) = 0 , 
\label{eqn:delta_phic} \\
&
\left[ \Box + \left( \bar{m}_{j}(T) \right)^2 + 2 ( 1 + 2\delta_{j0}) \lambda \bar{\phi}_{0c} \left( \delta \phi_{0c} \right)  \right] \phi_{js}
+ O\left(\left( \delta \phi_{0c} \right)^2 \right) = 0 , 
\label{eqn:delta_phis} \\
& \left( \bar{m}_j(T) \right)^2 = \lambda \left[ (1 + 2 \delta_{j0}) \left( \bar{\phi}_{0c} \right)^2 + (N+2) K_q(T) - v^2 \right] .
\end{align}
\end{subequations}

Finally, we derive the equation for $\phi_{js}$ by taking the solution of Eq.~\eqref{eqn:delta_phic} into account.
The solution of Eq.~\eqref{eqn:delta_phic}  is 
\begin{equation}
\delta \phi_{0c}   = -B \cos \left(\bar{m}_0(T) t + \theta \right) . 
\end{equation}
Substituting this solution into Eq.~\eqref{eqn:delta_phis}, 
and we obtain the following equation by changing of variable, $2\xi = \bar{m}_0(T) t + \theta$, 
and applying Fourier transformation:
\begin{equation} 
\left[ \frac{\partial^2}{\partial \xi^2} 
+ \frac{4 \left[\vec{k}^2 +  \left(\bar{m}_j(T) \right)^2 \right]}{\left(\bar{m}_0(T) \right)^2}
- \frac{ 8 \left( 1+2\delta_{j0} \right) \lambda \bar{\phi}_{0c} B}{\left(\bar{m}_0(T) \right)^2} \cos \left( 2 \xi \right)
\right] \phi_{js} \left(\xi,\vec{k}\right) = 0 .
\label{eqn:mathieu-expansionlesscase}
\end{equation} 
This equation is just a Mathieu equation. 
The amplified modes are derived from the above equation and are shown in the next section.


\subsection{Equation for soft mode in one dimensional expansion case}
The temperature decreases slowly at late time  in one dimensional expansion case.
Therefore, the motion of the condensate is quasi-periodic. 
In this subsection, we derive the equation for $\phi_{js}$ in one dimensionally expanding system.

The new variables, proper time $\tau$ and rapidity $\eta$,  are introduced:
\begin{subequations}
\begin{align}
& \tau = \sqrt{t^2 -z^2} , \\
& \eta = \frac{1}{2} \ln \left( \frac{t+z}{t-z} \right) .
\end{align}
\end{subequations}
The d'Alembertian is rewritten:
\begin{equation}
\Box = \frac{\partial^2}{\partial \tau^2} +  \frac{1}{\tau} \frac{\partial}{\partial \tau}  
            - \frac{1}{\tau^2} \frac{\partial^2}{\partial \eta^2}  - \frac{\partial^2}{\partial x^2} - \frac{\partial^2}{\partial y^2} .
\end{equation}
We assume that physical quantities are independent of $\eta$. 
The temperature decreases as 
\begin{equation}
T(\tau) = T_{\mathrm{ini}} \left( \frac{\tau_{\mathrm{ini}}}{\tau} \right)^{1/3}  ,
\end{equation}
where $T_{\mathrm{ini}}$ is the initial temperature. 

The equation of  $\delta\phi_{0c} = \phi_{0c} - \bar{\phi}_{0c}$ is derived by ignoring $O\left(\left( \delta \phi_{0c}\right)^2 \right)$:
The equation is 
\begin{equation}
\left[ \frac{\partial^2}{\partial \tau^2} +  \frac{1}{\tau} \frac{\partial}{\partial \tau}  +\left( \bar{m}_0(T) \right)^2 \right] (\delta \phi_{0c}) = 0 .
\end{equation}
The approximate solution for large $\tau$ is given by 
\begin{equation}
  (\delta \phi_{0c}) = -B \left( \frac{\tau_{\mathrm{ini}}}{\tau} \right)^{\frac{1}{2}} \cos \left( \bar{m}_0(T) \tau +\theta \right)  .
\end{equation}

The Fourier transformation of $\phi_{js}$ is applied to find amplified modes.
The field $\phi_{js}(\tau, \eta, k_{\perp})$ is decomposed as follows:
\begin{equation}
\phi_{js}(\tau, \eta, k_{\perp}) = \left( \frac{\tau_{\mathrm{ini}}}{\tau} \right)^{\frac{1}{2}} f_j(\tau, k_{\perp}) . 
\end{equation}
The equation of $f_j$ for large $\tau$ is given by 
\begin{align}
& 
\left[ \frac{\partial^2}{\partial \tau^2} + k_{\perp}^2 + \left( \bar{m}_{j}(T) \right)^2 \right.  
\nonumber \\ & \qquad \left.    
  - 2 (1+2\delta_{j0}) \lambda  \bar{\phi}_{0c} B \left( \frac{\tau_{\mathrm{ini}}}{\tau} \right)^{\frac{1}{2}} \cos \left( \bar{m}_0(T) \tau +\theta \right) 
  + O\left( (\delta \phi_{0c})^2 \right) \right] f_j = 0 .
\label{eqn:onedim:mathieu-like:base}
\end{align}
Temporarily, it is assumed that $T(\tau)$ is a constant to extract the amplified modes and that $\theta$ is set to zero. 
Applying the changing of variables $2 \xi_c =  \bar{m}_0(T) \tau$, 
we obtain
\begin{equation}
\left[ \frac{\partial^2}{\partial \xi_c^2} +  \frac{4 \left[ k_{\perp}^2 + \left( \bar{m}_{j}(T) \right)^2\right] }{ \left( \bar{m}_0(T) \right)^2} 
  - \frac{8 (1+2\delta_{j0}) \lambda \bar{\phi}_{0c} B }{ \left( \bar{m}_0(T) \right)^2}  \left( \frac{\xi_{c,\mathrm{ini}}}{\xi_c} \right)^{\frac{1}{2}} \cos \left( 2\xi_c \right) 
\right] f_j = 0 , 
\label{eqn:onedim:mathieu-like}
\end{equation}
where $\xi_{c, \mathrm{ini}}$ corresponds to $\tau_{\mathrm{ini}}$.
This equation is a Mathieu-like equation and  
the amplified modes are extracted approximately from Eq.~\eqref{eqn:onedim:mathieu-like}.


\section{Amplified modes}
\label{sec:amplified_modes}
In this section, the amplified modes are extracted analytically and numerically. 
The number of the fields $N$ is set to $4$. 
The parameters of the linear sigma model are set to $\lambda=20$,  $v=87.4 \mathrm{MeV}$, and $H=(119 \mathrm{MeV})^3$
\cite{Gavin1994,Ishihara2015,Ishihara1999}. 
At $T=0$, these parameters generate the sigma mass $m_0=  600 \mathrm{MeV}$,  the pion mass $m_j= 135 \mathrm{MeV}$ $(j \neq 0)$, 
and the pion decay constant $f_{\pi}= 92.5 \mathrm{MeV}$.

The Mathieu equation is characterized by two parameters $A$ and $Q$:
\begin{equation}
\left[ \frac{\partial^2}{\partial \xi^2} + A - 2Q \cos(2\xi) \right] g(\xi) = 0 .
\label{eqn:mathieu}
\end{equation}
The amplification occurs around $A=n^2$ when $Q \neq 0$, where $n$ is a positive integer. 
The amplified modes are extracted by comparing the equation of motion to the Mathieu equation.

\subsection{Amplified modes in an expansionless case}
In the case of an expansionless case,  the amplified mode is easily extracted 
by comparing  Eq.~\eqref{eqn:mathieu-expansionlesscase} with Eq.~\eqref{eqn:mathieu}. 
The parameter $A$ for the sigma field, $A^{\sigma}$,  and that for the pion fields, $A^{\pi}$, are given by 
\begin{subequations} 
\begin{align}
A^{\sigma} &=
\frac{4 \left[\vec{k}^2 +  \left(\bar{m}_0(T) \right)^2 \right]}{\left(\bar{m}_0(T) \right)^2} ,
\label{A:sigma}
\\
A^{\pi} & =
\frac{4 \left[\vec{k}^2 +  \left(\bar{m}_j(T) \right)^2 \right]}{\left(\bar{m}_0(T) \right)^2} \qquad (j \neq 0) , 
\label{A:pi}
\end{align}
\label{expansionless:A}
\end{subequations} 
where the suffix $j$ for pion fields is omitted.
The zero mode is not the candidate of the amplified mode, because  the condensate corresponds to the zero mode. 
Therefore, the amplified modes for the sigma field are given by the equation $A^{\sigma} = n^2$ ($n \ge 3$). 
The amplified modes for the pion fields are given by the equation $A^{\pi} = n^2$.
The finite modes corresponding to $n \ge 2$ exist, because the pion mass is lighter than the sigma mass. 
The existence of the mode corresponding to $n=1$ depends on the parameters of the linear sigma model.

The magnitudes of the amplified modes are given by 
\begin{subequations}
\begin{align}
|\vec{k}^{\sigma}| &= \Bigg( \sqrt{\frac{n^2}{4} -1} \Bigg) \left(\bar{m}_0(T) \right) ,  
\label{expansionless:amplifiedmode:sigma}
\\
|\vec{k}^{\pi}| &=  \Bigg( \sqrt{\frac{n^2}{4}   -  \frac{\left(\bar{m}_j(T) \right)^2}{\left(\bar{m}_0(T) \right)^2}} \Bigg) \left(\bar{m}_0(T) \right)  
\qquad (j \neq 0) , 
\label{expansionless:amplifiedmode:pi}
\end{align}
\end{subequations}
where $\vec{k}^{\sigma}$ and $\vec{k}^{\pi}$ are the amplified mode for the sigma field and that for pion field, respectively.
A positive integer $n$ for the pion field is realized  
when $n$ satisfies the condition, $n^2  \left(\bar{m}_0(T) \right)^2 /4 -  \left(\bar{m}_j(T) \right)^2 > 0$.

\begin{figure}
\begin{center}
\includegraphics[width=0.45\textwidth]{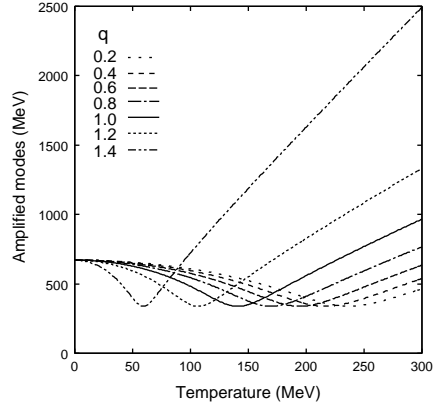}
\caption{Amplified modes for the sigma field at $n=3$ for various $q$}
\label{fig:expansionless:sigma}
\end{center}
\end{figure}

Figure~\ref{fig:expansionless:sigma} shows the amplified modes of the sigma field at $n=3$ for various $q$.
The temperature dependences of the amplified modes reflect directly the temperature dependence of the sigma mass.
The distribution has a long tail for $q>1$,  and the expectation value of $\left(\phi_{ih}\right)^2$ at $q>1$ is larger than that at $q=1$. 
Therefore, the condensate at $q>1$ is smaller than that at $q=1$, 
and  the temperature at which the sigma mass at $q>1$ reaches the minimum is lower than that at which the sigma mass at $q=1$.
As a result, the amplified mode as a function of the temperature behaves like the figure.

\begin{figure*}
\begin{center}
\includegraphics[width=0.45\textwidth]{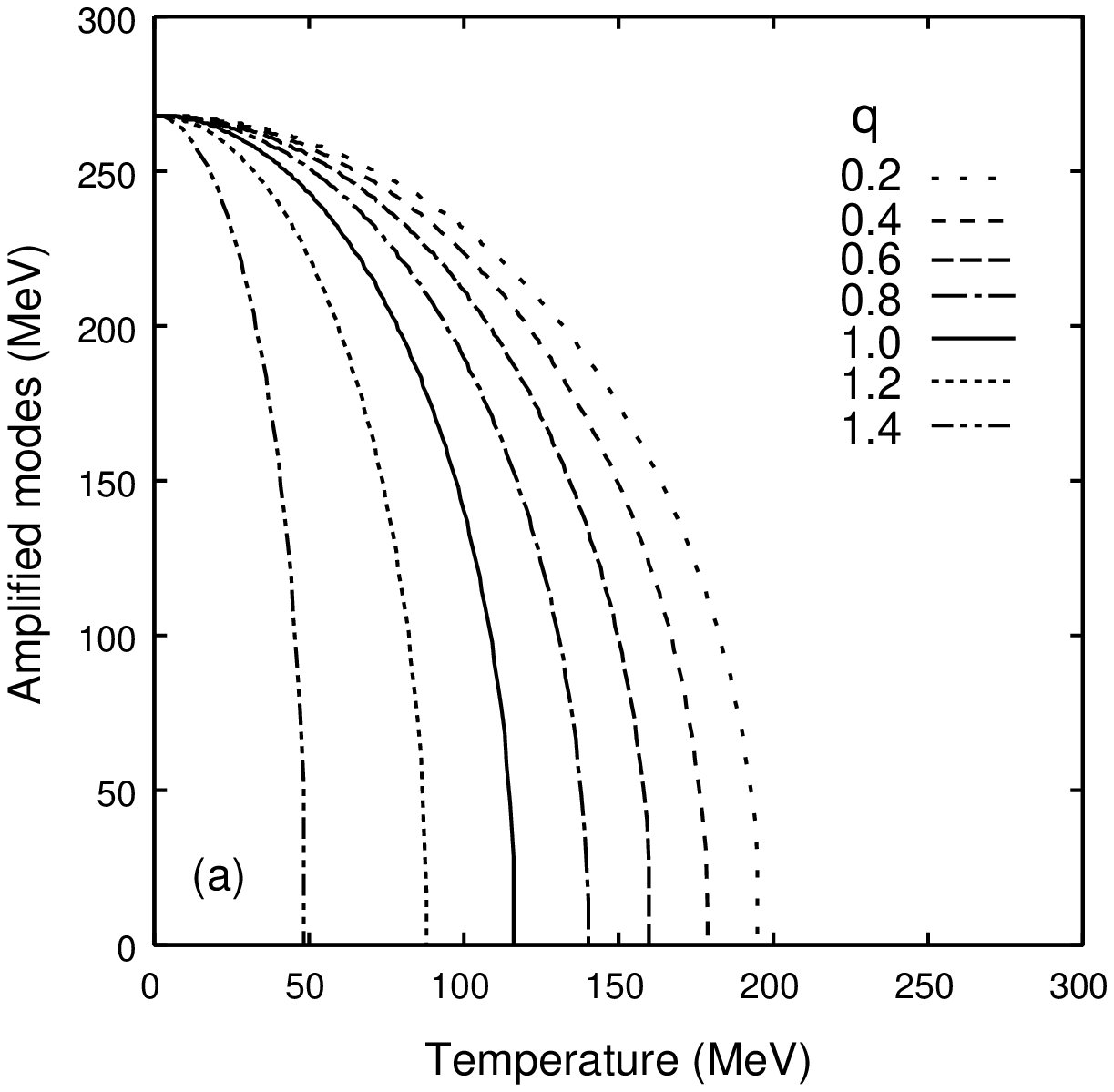}
\includegraphics[width=0.45\textwidth]{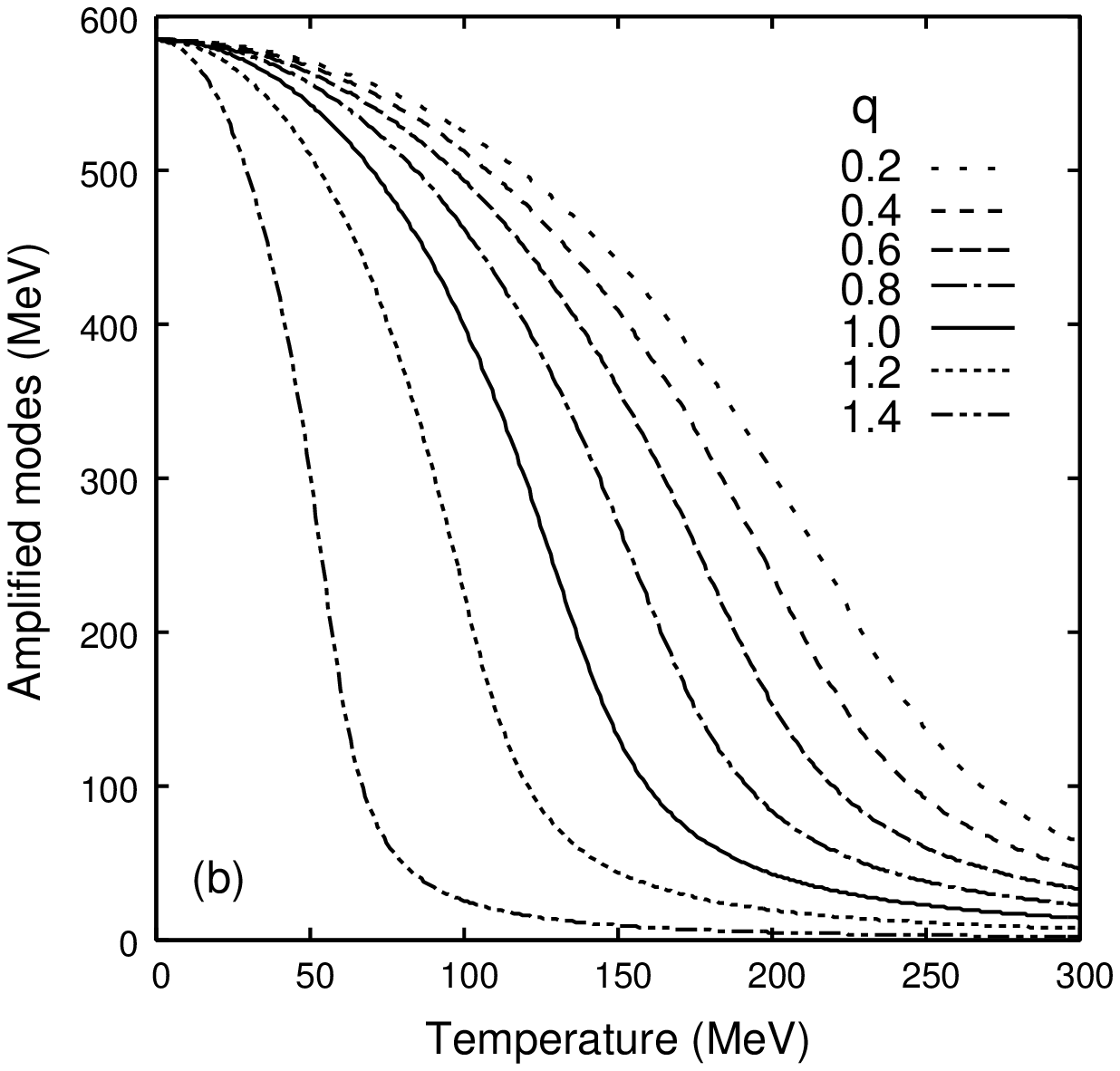}
\includegraphics[width=0.45\textwidth]{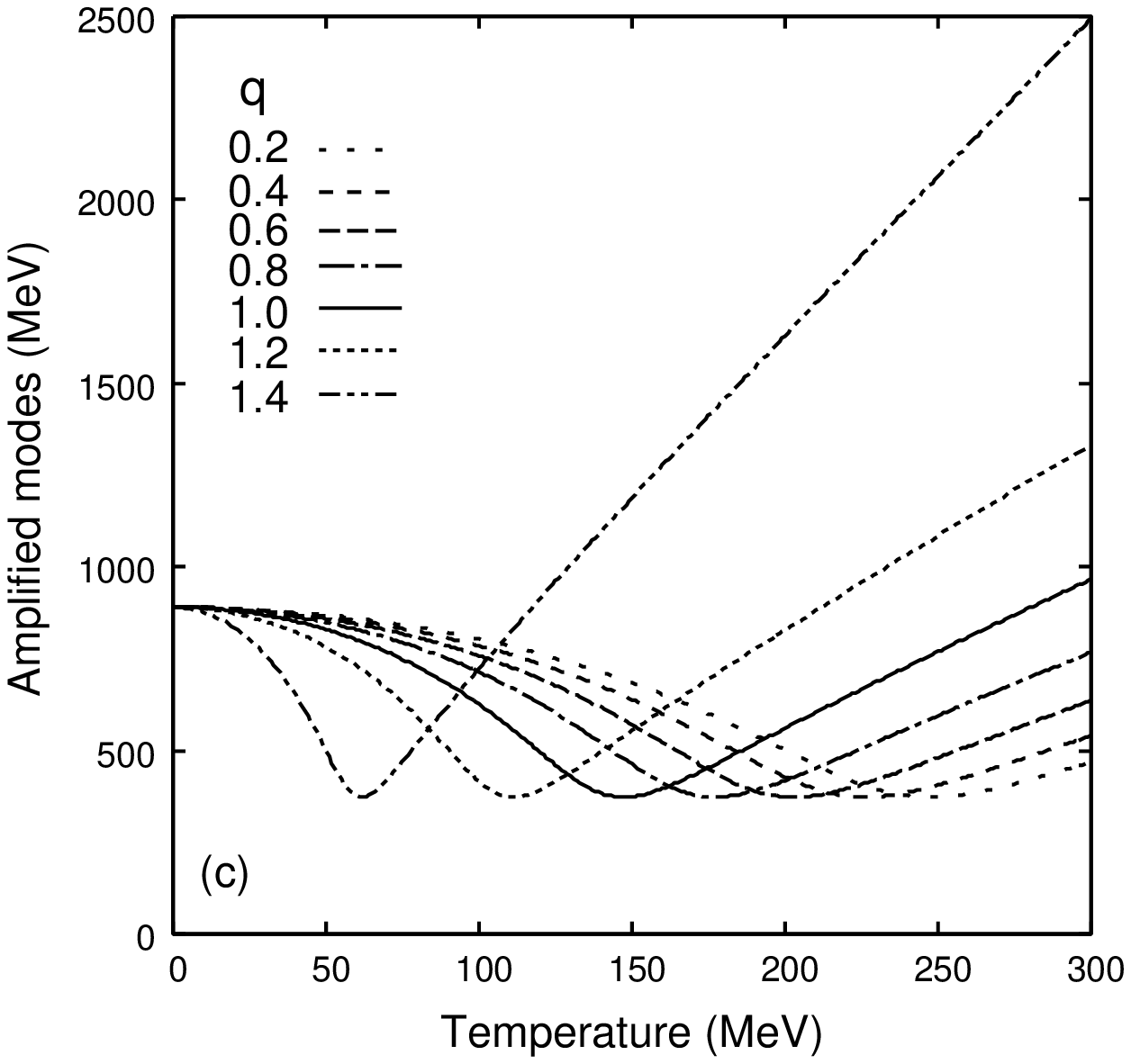}
\caption{Amplified modes for the pion fields for various $q$: 
(a) modes at $n=1$, (b) modes at $n=2$, and (c) modes at $n=3$. 
}
\label{fig:expansionless:pion}
\end{center}
\end{figure*}

Figure~\ref{fig:expansionless:pion} shows the amplified modes for the pion fields for various $q$.
Figure~\ref{fig:expansionless:pion}(a) is the modes at $n=1$, (b) is at $n=2$, and (c) is at $n=3$. 
The difference between the pion mass and the sigma mass becomes small as the temperature increases. 
Therefore, as shown in Fig.~\ref{fig:expansionless:pion}(a), 
the amplified mode at $n=1$ becomes smaller as the temperature increases, and the mode vanishes. 
This implies that the resonance band at $n=1$ vanishes.  
The amplified mode at $n=2$ goes to zero as the temperature increases,  
because $\bar{m}_j(T) /\bar{m}_0(T) $ approaches to $1$ as the temperature increases. 
In contrast, the amplified modes at $n \ge 3$ becomes large at high temperature, as shown in Fig.~\ref{fig:expansionless:pion}(c). 
This behavior comes from the temperature dependence of the sigma mass.  
The magnitude of the amplified mode of the first resonance bands decreases as $q$ increases, as shown in Fig.~\ref{fig:expansionless:pion}(a).
This behavior is also shown in Fig.~\ref{fig:expansionless:pion}(b).
The behavior comes from the fact that the tail of the distribution becomes long as $q$ increases. 

\subsection{Amplified modes in one dimensional expansion case}
In one dimensional expansion case,  the temperature $T(\tau)$ decreases slowly. 
The approximate amplified modes are obtained  from Eq.~\eqref{eqn:onedim:mathieu-like}.
The parameter $A$ is obtained by replacing $\vec{k}$ by $\vec{k}_{\perp}$ in Eq.~\eqref{expansionless:A}.
Therefore, the amplified mode for the sigma field, $\vec{k}_{\perp}^{\sigma}$, is obtained by replacing  $\vec{k}^{\sigma}$ by $k_{\perp}^{\sigma}$
in Eq.~\eqref{expansionless:amplifiedmode:sigma}, 
and the amplified mode for the pion field, $\vec{k}_{\perp}^{\pi}$, is obtained by replacing  $\vec{k}^{\pi}$ by $\vec{k}_{\perp}^{\pi}$ 
in Eq.~\eqref{expansionless:amplifiedmode:pi}:
\begin{subequations}
\begin{align}
|\vec{k}_{\perp}^{\sigma}| &= \Bigg( \sqrt{\frac{n^2}{4} -1} \Bigg) \left(\bar{m}_0(T) \right) \qquad  (n = 3, 4, \cdots) , 
\label{onedim:amplifiedmode:sigma}
\\
|\vec{k}_{\perp}^{\pi}| &=  \Bigg( \sqrt{\frac{n^2}{4}   -  \frac{\left(\bar{m}_j(T) \right)^2}{\left(\bar{m}_0(T) \right)^2}} \Bigg) \left(\bar{m}_0(T) \right) . 
\label{onedim:amplifiedmode:pi}
\end{align}
\end{subequations}
The right-hand sides of the above equations are identical 
to those of Eqs.~\eqref{expansionless:amplifiedmode:sigma}  and \eqref{expansionless:amplifiedmode:pi}.
The amplified modes extracted from the above equations are the same.
Realistically, the amplified modes shift as the time $\tau$ increases, because the temperature decreases as the time increases.
Therefore, the numerical calculations are required to find the amplified modes. 

It is better to use dimensionless variables in numerical calculations. 
The transformation $2 \xi_c =  \bar{m}_0(T) \tau$ is not valid, because $T$ is time-varying. 
Instead, we set $\theta=0$ and apply the changing of variable $2 \xi_s =  \bar{m}_0(T=0) \tau$, 
and we obtain the following equation from Eq.~\eqref{eqn:onedim:mathieu-like:base} for the numerical studies:
\begin{align}
& 
\left[ \frac{\partial^2}{\partial \xi_s^2} + \frac{4\left[ k_{\perp}^2 + \left( \bar{m}_{j}(T) \right)^2 \right]}{ \left( \bar{m}_0(T=0) \right)^2} 
\right. \nonumber \\ & \qquad  \left. 
  - \frac{8 (1+2\delta_{j0}) \lambda \bar{\phi}_{0c} B }{ \left( \bar{m}_0(T=0) \right)^2}  \left( \frac{\xi_{\mathrm{s, ini}}}{\xi_{\mathrm{s}}} \right)^{\frac{1}{2}} 
\cos \left( 2 \left( \frac{\bar{m}_0(T)}{\bar{m}_{0}(T=0)} \right) \xi_s \right) 
\right] f_j = 0 .
\label{eqn:numerical_eq}
\end{align}
This Mathieu-like equation was used to calculate the quantities numerically 
with the initial time $\xi_{\mathrm{s},\mathrm{ini}}$, the initial temperature $T_{\mathrm{ini}}$, and the amplitude $B$.

The magnitude of the amplification was estimated as follows. 
The quantity  $f_{j}$ was calculated numerically from $\xi_{\mathrm{s}}=\xi_{\mathrm{s},\mathrm{ini}}$ to $\xi_{\mathrm{s}}=10000$ 
with $f_{j}(\xi_{\mathrm{s}} =\xi_{\mathrm{s}, \mathrm{ini}}) =1$ and  $df_{j}(\xi_{\mathrm{s}})/d\xi_{\mathrm{s}} =0$ at $\xi_{\mathrm{s}} = \xi_{\mathrm{s},\mathrm{ini}}$.
The local extremums  of $f_j$ were extracted, 
and the extremums of $f_j$ in the region of $[9500:10000]$ were fitted with a constant function.
The value of the constant function was regarded as the magnitude of the amplification of $f_j$.

\begin{figure}
\begin{center}
\includegraphics[width=0.5\textwidth]{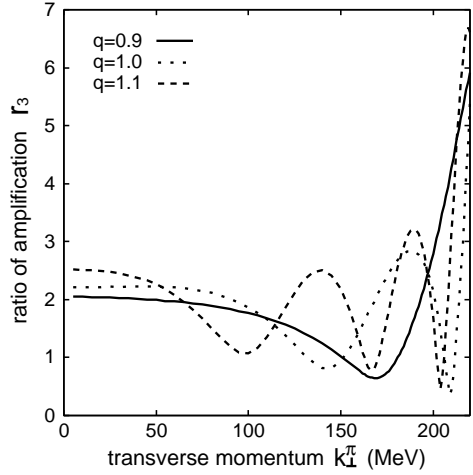}
\end{center}
\caption{Amplification of $f_{3}$ for various $q$. 
The quantity $r_3(k_{\perp}, q)$ is the ratio of the amplitude of $f_{3}(\xi_s=10000, k_{\perp})$ to the amplitude $f_{3}(\xi_{\mathrm{ini}}, k_{\perp})$.
The range of $k_{\perp}^{\pi}$ (MeV) in this figure is $[5,220]$. 
The initial time $\xi_{s, \mathrm{ini}}$, the initial temperature $T_{\mathrm{ini}}$, and the amplitude $B$ are 15, 160 MeV, and 10 MeV, respectively.}
\label{fig:amlifcation_for_pi_field}
\end{figure}

Figure~\ref{fig:amlifcation_for_pi_field} shows the amplification of $f_3$ 
for various $q$ in the range of 5MeV $\le k_{\perp}^{\pi} \le$  220MeV. 
The initial time $\xi_{\mathrm{s}, \mathrm{ini}}$, the initial temperature $T_{\mathrm{ini}} \equiv T(\xi_{\mathrm{s}, \mathrm{ini}})$, 
and the amplitude $B$ were set to $15$, $160\mathrm{MeV}$, and $10$MeV, respectively. 
We define the quantity $r_j(k_{\perp}, q)$ as the ratio of the amplitude of $f_{j}(\xi_s=\infty, k_{\perp})$ to the amplitude $f_{j}(\xi_{\mathrm{ini}}, k_{\perp})$.
In numerical calculations, we replace the amplitude of $f_{j}(\xi_s=\infty, k_{\perp})$ by the amplitude of $f_{j}(\xi_s=10000, k_{\perp})$.
The amplitude is remarkably large around $k_{\perp}^{\pi} = 268$ MeV, 
and it is difficult to depict the figure around $k_{\perp}^{\pi} = 268$ MeV. 
Therefore, the range of $k_{\perp}^{\pi}$ in the figure is $[5, 220]$.
The magnitude of the amplification in this figure oscillates. 
The amplitudes of the oscillations become large as $k_{\perp}^{\pi}$ increases.
After that, the amplitudes decrease and the amplification is weak in the range of 270 MeV $\le k_{\perp}^{\pi} \le$ 300 MeV. 
In Fig.~\ref{fig:amlifcation_for_pi_field}, the amplification occurs even for small $k_{\perp}^{\pi}$ and is strong for large $q$. 

\begin{figure}
\begin{center}
\includegraphics[width=0.5\textwidth]{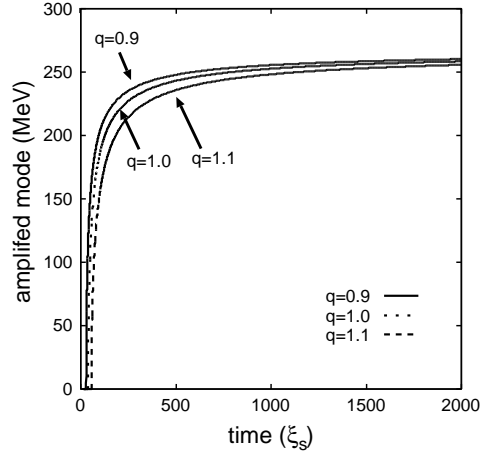}
\end{center}
\caption{Time development of amplification mode of $f_{3}$, Eq.~\eqref{onedim:amplifiedmode:pi}, 
at $n=1$ for various $q$, with $\xi_{\mathrm{s}, \mathrm{ini}} =15$ and $T_m =$160 MeV.
The initial time $\xi_{\mathrm{s}, \mathrm{ini}}$ and the temperature $T_m$ are 15 and 160 MeV, respectively. 
}
\label{fig:amlifed_mode}
\end{figure}

The value $r_{3}(k_{\perp}^{\pi}, q)$ as a function of $k_{\perp}^{\pi}$ grows as  $k_{\perp}^{\pi}$ increases in Fig.~\ref{fig:amlifcation_for_pi_field}.
This growth can be explained by the existence of the first resonance band determined from Eq.~\eqref{onedim:amplifiedmode:pi}. 
The amplified mode in the amplified region of $n=1$ at $T=0$ is 268 MeV approximately. 
The amplified mode varies as $T$ increases realistically, 
and the amplified mode converges to 268 MeV as $\xi_s$ increases. 
Figure~\ref{fig:amlifed_mode} shows the amplified modes, Eq.~\eqref{onedim:amplifiedmode:pi}, at $n=1$ for various $q$.
It is easily shown that the amplified modes converge to 268 MeV and  
that the mode at small $q$ is larger than that at large $q$.
The amplified mode of $f_3$ at $n=1$ is small at the beginning of the expansion, as shown in Fig.~\ref{fig:amlifed_mode}. 
The mode increases,  and reaches the asymptotic value.
This behavior indicates that $f_3$ with small $k_{\perp}^{\pi}$ grows in the early stage of the expansion and 
that the amplifications for soft modes occur. 
The amplification is not always weak, 
because the coefficient of the oscillating term of Eq.~\eqref{eqn:numerical_eq} decreases slowly as the time $\tau$ increases. 
This feature can be seen by estimating the exponent roughly.
We express the function $f_j$ in Eq.~\eqref{eqn:onedim:mathieu-like:base}
as ${\displaystyle f_j =C \exp \left( \int^{\tau} du \ s(u) \right)}$, where $C$ is constant.
The function diverges when the quantity ${\displaystyle \left( \int^{\tau}  du\ s(u) \right)}$ diverges.  
It is shown that the quantity becomes large or diverges from the rough estimation,
as shown in \ref{sec:evaluation_of_exponent_one_dim}. 
This fact implies that ${\displaystyle f_j}$ in Eq.~\eqref{eqn:onedim:mathieu-like:base} is quite large or divergent
around the amplified mode at $n=1$.


The amplification for small $k_{\perp}^{\pi}$  can be seen in one dimensional expansion case. 
The trajectories of the coefficients of the Mathieu-like equation are helpful to understand the amplification.
The following quantities are introduced to make the amplification of $f_3$ clear:
\begin{subequations}
\begin{align}
\tilde{A}^{\pi} &:= 4 \left[ \frac{k_{\perp}^2 + \left( \bar{m}_{3}(T) \right)^2 }{ \left( \bar{m}_0(T) \right)^2} \right] , 
\label{def:onedim:Api} \\
\tilde{Q}^{\pi} &:= \frac{4\bar{\phi}_{0c} \lambda B }{ \left( \bar{m}_0(T) \right)^2}  \left( \frac{\xi_{\mathrm{c}, \mathrm{ini}}}{\xi_{\mathrm{c}}} \right)^{\frac{1}{2}}
= \frac{4\bar{\phi}_{0c} \lambda B }{ \left( \bar{m}_0(T) \right)^2}  \left( \frac{\bar{m}_{0}(T_{\mathrm{ini}}) \xi_{\mathrm{s}, \mathrm{ini}}}{\bar{m}_{0}(T) \xi_{\mathrm{s}}} \right)^{\frac{1}{2}} . 
\label{def:onedim:Qpi}
\end{align}
\label{def:onedim:Api:Qpi}
\end{subequations}
The quantity $\left( \xi_{\mathrm{c, \mathrm{ini}}}/{\xi_{\mathrm{c}}} \right)$ is equal to the quantity $\left( {\xi_{\mathrm{s}, \mathrm{ini}}}/{\xi_{\mathrm{s}}} \right)$ 
when $\bar{m}_0(T)$ is constant. 
Equation~\eqref{eqn:onedim:mathieu-like:base} is rewritten:
\begin{equation}
\left[ \frac{\partial^2}{\partial \xi_{\mathrm{c}}^2} + \tilde{A}^{\pi}  - 2\tilde{Q}^{\pi} \cos \left( 2\xi_{\mathrm{c}} \right)  \right] f_3 = 0 .
\label{eqn:coef:replaced}
\end{equation}
Therefore, the amplification of $f_3$ can be understood by drawing the trajectory of $(\tilde{Q}^{\pi}, \tilde{A}^{\pi})$. 
We note again that the above Eq.~\eqref{eqn:coef:replaced} was derived when the temperature is slowly varying  
and 
that the condition, the temperature is slowly varying, is not assumed in numerical calculations with Eq.~\eqref{eqn:numerical_eq}.
Parametric amplification can be discussed by studying the time evolutions of $(\tilde{Q}^{\pi}, \tilde{A}^{\pi})$ and their trajectories.

\begin{figure}
\begin{center}
\includegraphics[width=0.5\textwidth]{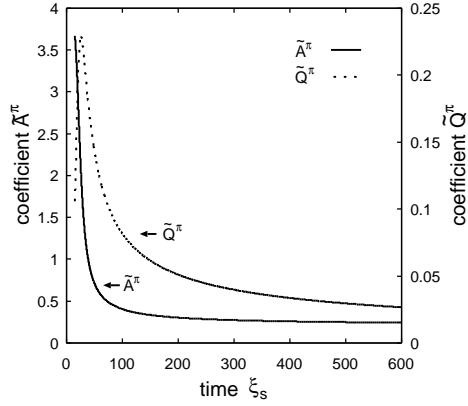}
\end{center}
\caption{Time developments of $\tilde{A}^{\pi}$  and $\tilde{Q}^{\pi}$ with $k_{\perp}^{\pi}=5$ MeV for $f_3$.
The initial time  $\xi_{\mathrm{s}, \mathrm{ini}}$, initial temperature $T_{\mathrm{ini}}$, and parameter $B$ 
are 15, 160 MeV, and 10 MeV, respectively.}
\label{fig:time_development:A:Q}
\end{figure}

Figure~\ref{fig:time_development:A:Q} shows the time developments of $\tilde{A}^{\pi}$  and $\tilde{Q}^{\pi}$ with $k_{\perp}^{\pi}=5$ MeV.
The coefficients, $\tilde{A}^{\pi}$ and  $\tilde{Q}^{\pi}$, change in the early time of the evolution,  and approach to the asymptotic values, respectively.
The coefficient $\tilde{Q}^{\pi}$ becomes large temporarily.
This variation of $\tilde{Q}^{\pi}$ comes from the variation of $\bar{m}_0(T)$ that has a minimum at a certain temperature\cite{Ishihara2015}. 
The position of $(\tilde{Q}^{\pi}, \tilde{A}^{\pi})$ on the $\tilde{Q}^{\pi}$-$\tilde{A}^{\pi}$ plane is obtained by plotting these values. 

\begin{figure}
\begin{center}
\includegraphics[width=0.5\textwidth]{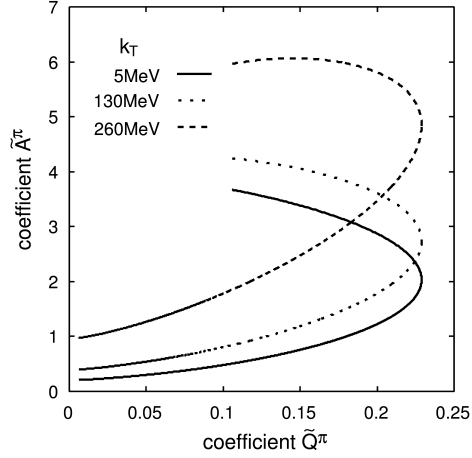}
\end{center}
\caption{
The trajectories on the $\tilde{Q}^{\pi}$ -$\tilde{A}^{\pi}$ plane at $q=1.0$ for $k_{\perp}^{\pi}=5$ MeV, $130$ MeV and $260$ MeV.
The initial time  $\xi_{\mathrm{s}, \mathrm{ini}}$, initial temperature $T_{\mathrm{ini}}$, and the parameter $B$ are 15,  160 MeV, and 10 MeV, respectively.
}
\label{fig:Q-A-plane:kdep}
\end{figure}

Figure~\ref{fig:Q-A-plane:kdep} shows the trajectories of $(\tilde{Q}^{\pi}, \tilde{A}^{\pi})$ on the $\tilde{Q}^{\pi}$-$\tilde{A}^{\pi}$ plane
at $q=1.0$ for $k_{\perp}^{\pi} = 5$ MeV, $130$ MeV, and $260$ MeV.
As seen in Eqs.~\eqref{def:onedim:Api} and \eqref{def:onedim:Qpi},
the quantity $\tilde{Q}^{\pi}$ does not depend on $k_{\perp}^{\pi}$, and the difference of the trajectories depends on $\tilde{A}^{\pi}$.
As is shown in Fig.~\ref{fig:Q-A-plane:kdep}, $\tilde{A}^{\pi}$ at $k_{\perp}^{\pi}=5$ MeV is approximately 4 at the initial time,
because $\bar{m}_{j}$ $(j=1,2,3)$  is close to $\bar{m}_{0}$ at high temperature.
The points $(\tilde{Q}^{\pi}, \tilde{A}^{\pi})$ for $k_{\perp}=5$ MeV and $k_{\perp}=130$ MeV move in the first resonance band, and move out after that. 
It is possible for the field to be amplified when the point $(\tilde{Q}^{\pi}, \tilde{A}^{\pi})$ is on the resonance band for a long time.
The point $(\tilde{Q}^{\pi}, \tilde{A}^{\pi})$ for $k_{\perp} \sim 268$ MeV stays on the first resonance band for a long time, 
and the field of $k_{\perp} \sim268$ MeV is amplified.

\begin{figure}
\begin{center}
\includegraphics[width=0.5\textwidth]{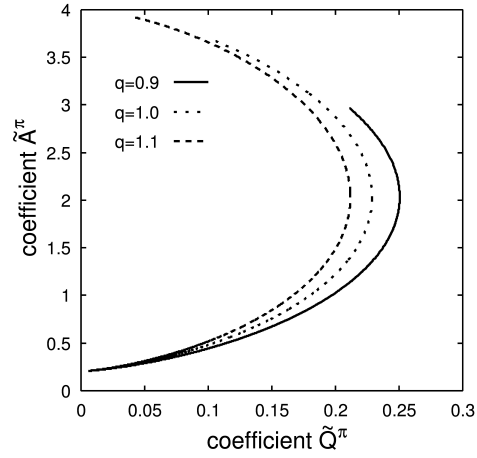}
\end{center}
\caption{
The trajectories on the $\tilde{Q}^{\pi}$ -$\tilde{A}^{\pi}$ plane at $k_{\perp}^{\pi}=5$ MeV  for $q=0.9$, $1.0$, and $1.1$. 
The initial time  $\xi_{\mathrm{s}, \mathrm{ini}}$, initial temperature $T_{\mathrm{ini}}$, and the parameter $B$ are 15,  160 MeV, and 10 MeV, respectively.
}
\label{fig:Q-A-plane:qdep}
\end{figure}

The motions of $(\tilde{Q}^{\pi}, \tilde{A}^{\pi})$ for $q=0.9$ and $1.1$ are similar.
Figure~\ref{fig:Q-A-plane:qdep} shows the trajectories of $(\tilde{Q}^{\pi}, \tilde{A}^{\pi})$ at $k_{\perp}^{\pi} = 5$ MeV 
for $q=0.9$, $1.0$, and $1.1$ on the $\tilde{Q}^{\pi}$-$\tilde{A}^{\pi}$ plane. 
The curves in the figure are similar. 
The amplification at $q=0.9$ is weakest in the low $k_{\perp}$ region in Fig.~\ref{fig:amlifcation_for_pi_field}
though the coefficient $\tilde{Q}^{\pi}$ at $q=0.9$ is largest.
\section{Discussion and Conclusion}
\label{sec:discussion_and_conclusion}
We studied the effects of Tsallis distribution on parametric resonance in chiral phase transitions. 
We used the linear sigma model and investigated the amplification of soft modes caused by the motion of the condensate
under the assumption that the distribution of hard modes is described by a Tsallis distribution. 
We treated expansionless and one dimensional expansion cases and did not use the expectation value used in the Tsallis statistics in this study:  
The standard expectation value was used. 
A Tsallis distribution has two parameters.  
The parameters $T$ and $q$ are called temperature and entropic parameter respectively in this paper, 
because the parameter $T$ is the temperature of the Boltzmann-Gibbs statistics at $q=1$. 
We note here that the value of $q$ is restricted from the energetic point of view \cite{Ishihara2015}, 
though we showed the results at $q=1.4$ in the expansionless case.

The amplification for the soft mode was found in the expansionless case.
The temperature is fixed, and the motion of the soft mode is described approximately by a Mathieu equation. 
The amplified modes are determined by the Mathieu equation.
The lowest amplified mode of the sigma field is the mode on the third resonance band ($n=3$). 
The amplified mode of the sigma field has a minimum as a function of the temperature. 
The larger the value of $q$ is, the lower the temperature at the minimum is. 
This comes from the temperature dependence of the sigma mass which reflects the fluctuations of the fields. 
The magnitude of the amplified mode decreases, reaches the minimum, and increases after that. 
The lowest amplified mode of the pion field is the mode on the first resonance band ($n=1$) at low temperature 
and the mode on the second resonance band ($n=2$) at high temperature. 
The amplified mode at $n=1$ decreases and vanishes as the temperature increases,  
because the difference in mass between sigma meson and pion decreases as the temperature increases. 
The amplified mode at $n=2$ also decreases as the temperature increases.  
The mode at $n=2$ exists even when the temperature is high. 
The temperature dependence of the mode at $n \ge 3$ for the pion field is similar to that for the sigma field.
The $q$ dependence of the amplified mode comes from the $q$ dependences of the masses. 
As shown in the numerical results of the pion field, 
the amplified mode for small $q$ decreases slowly as a function of $T$, while the mode for large $q$ decreases rapidly. 

The amplified mode is soft for the pion field when the temperature is appropriate,
because the amplified modes for the first and second resonance bands decrease as the temperature increases. 
Contrarily, the mode for the sigma field and the mode at $n \ge 3$ for the pion field cannot be soft enough, 
because the amplified modes are lower bounded.

The amplification for the pion field was found by removing the effects of the expansion as 
$\phi_j(\tau, k_{\perp}) = \left(\tau_{\mathrm{ini}}/\tau\right)^{1/2} f_{j}(\tau, k_{\perp})$ in one dimensional expansion case  
under the assumption that the quantities are independent of the rapidity $\eta$,  
where $\tau$ is the proper time, $\tau_{\mathrm{ini}}$ is the initial proper time, and $k_{\perp}$ is the transverse momentum.
The ratio of the amplitude of $f_{3}(\tau=\infty, k_{\perp}^{\pi})$ to the initial value $f_{3}(\tau_{\mathrm{ini}}, k_{\perp}^{\pi})$,  
$r_3(k_{\perp}^{\pi}, q)$, was studied and the following facts were found, where $k_{\perp}^{\pi}$ is the transverse momentum of the pion field:
1)  $r_3(k_{\perp}^{\pi}, q)$ for soft mode is larger than $1$, 
2)  $r_3(k_{\perp}^{\pi}, q)$ as a function of $k_{\perp}^{\pi}$ oscillates around the amplified mode of the first resonance band, 
and 
3) the amplitude of $r_3(k_{\perp}^{\pi}, q)$ as a function of $k_{\perp}^{\pi}$ is extremely large around the amplified mode of the first resonance band. 


In summary, 
the equation of soft mode is described by a Mathieu equation in an expansionless case. 
The magnitudes of the amplified modes decrease as $q$ increases.
The larger the value $q$ is,  the softer the mode is.  
For the pion fields, the amplified modes for the first and second resonance bands decrease as the temperature $T$ increases. 
The mode of the first resonance band vanishes at a certain temperature. 
The larger the value $q$ is, and lower the temperature is.
The mode of the second resonance band decreases and approaches to zero. 
The amplified mode of the second resonance band remains at high temperature. 

The equation of soft mode is described by a Mathieu-like equation in one dimensional expansion case.
The soft modes are amplified for the pion fields.
The strong amplification can be seen around the amplified mode of the first resonance band of the Mathieu equation at $T=0$, 
and the magnitude of the amplification around the amplified mode of the first resonance band at $T=0$ varies frequently.

The parametric resonance occurs for the pion fields in both the cases.
In the expansionless case, the field on the resonance band is amplified.
In one dimensional expansion case, the soft modes are amplified 
and the amplification is strong around the amplified mode of the first resonance band of the Mathieu-like equation at $T=0$.

We hope that this work will be helpful for the recognition of the amplification in high energy collisions.




\appendix
\section{The evaluation of the amplification in one dimensional expansion case}
\label{sec:evaluation_of_exponent_one_dim}
In this appendix, we evaluate the amplification of the field approximately in one dimensional expansion case.  
The following equation is used to evaluate the amplification according to Ref.~\refcite{ishihara2004}.
\begin{equation}
\left[ \partial_t^2 + \omega^2(t) + p(t) \cos \left(\Omega(T) t \right) +  q(t) \cos \left(2 \Omega(T) t \right) \right] \xi(t) = 0 .
\label{eqn:base_diff_eq}
\end{equation}
The solution $\xi(t)$ is described by the following expansion. 
\begin{equation}
\xi(t) = \sum_{n=1} \left[P_n(t) \exp \left( -i n \Omega(t) t /2 \right) + P^{*}_n(t) \exp \left(i n \Omega(t) t /2 \right)  \right] + R(t).
\end{equation}
We focus on the amplification of $n=1$. 
The amplification is evaluated approximately by putting $P_3=0$ for $n=1$.  
The new function $\alpha_1(t)$ is introduced as follows:
\begin{equation}
\alpha_1(t) := \left( P_1(t)+P_1^*(t) \right)/2 = \alpha_{1c} \exp \left( \int^t \ du \ s_1(u) \right) ,
\end{equation}
where $\alpha_{1c}$ is constant.
The approximate expression of $s_1(t)$ is obtained: 
\begin{equation}
\left[s_{1}(t)\right]^2 = \left[\Omega(t)\right]^{-2} \left\{ \frac{1}{4} \left[ p(t)\right]^2 - \left[ \omega^2(t) - \frac{1}{4} \Omega^2(t) \right]^2 \right\} . 
\label{eqn:base_equation}
\end{equation}

To find the expression of $s_1(t)$ in one dimensional expansion case, 
the function $p(t)$ is set to $p_0/\sqrt{t}$.    
The quantity $s_1(t)$ is estimated 
when $\Omega(t)$ and $\omega(t)$ are approximately equal to $\Omega(\infty) $ and $\omega(\infty)$ respectively at large $t$:
The quantity $s_1(t)$ is approximately evaluated with Eq.~\eqref{eqn:base_equation}:
\begin{equation}
\left[s_{1}(t)\right]^2 \sim 
\left[\Omega(\infty)\right]^{-2} \left\{ \frac{1}{4}  \frac{\left(p_0\right)^2}{t}  - \delta^2 \right\} , 
\end{equation}
where quantity $\delta$ is defined by 
\begin{equation}
\delta :=  \omega^2(\infty) - \frac{1}{4} \Omega^2(\infty).
\end{equation}
The field is amplified when $s_1(t)$ is larger than $0$,  
and the amplification occurs for $t < t^*$,  where $t^*$ is defined as  
${\displaystyle t^* := {\left(p_0\right)^2 }/{(4\delta^2)}  }$.

The amplification from time $t_0$ to $t=\infty$ is evaluated by  the following quantity:
\begin{equation}
\int^{\infty}_{t_0} \ du \ s_1(u) \ \Theta(s_1(u)) \sim \frac{1}{\Omega(\infty)} \int_{t_0}^{t^*} \ du \ \sqrt{\frac{\left( p_0 \right)^2}{4u} - \delta^2},  
\label{eqn:s1:evaluation}
\end{equation}
where $t_0$ should be large enough to hold the relations, $\Omega(t) \sim \Omega(\infty) $ and $\omega(t) \sim \omega(\infty)$,  
and $\Theta(x)$ is the step function which is $1$ for $x>0$ and $0$ for $x<0$.
The solution of Eq.~\eqref{eqn:base_diff_eq} should be calculated for a long time in the case of small $\delta$ 
when the amplification is evaluated numerically. 

The amplification is strongest when $\delta=0$. 
The left-hand side of Eq.~\eqref{eqn:s1:evaluation} for $\delta =0$ is 
\begin{equation}
\int^{\infty}_{t_0} \ du \ \left. s_{1}(u) \right|_{\delta=0}  \ \Theta\left(  \left. s_1(u) \right|_{\delta=0} \right) 
\sim 
\frac{|p_0|}{\Omega(\infty)} \ \left[ u^{1/2}  - u_0^{1/2}  \right]_{t_0}^{\infty}
\end{equation}
Therefore, the integral is 
\begin{equation}
\int^{\infty}_{t_0} \ du \ \left. s_{1}(u) \right|_{\delta=0} = \infty . 
\end{equation}

This implies that the magnitude of the field is quite large or diverges as $t$ approaches to $\infty$. 
Therefore,  the numerical calculation around $\delta=0$ should be performed for a long time 
to evaluate the amplification.

\end{document}